# New Constructions of a Family of 2-Generator Quasi-Cyclic Two-Weight Codes and Related Codes

Eric Zhi Chen
School of Engineering
Kristianstad University
291 88 Kristianstad
Sweden
eric.chen@tec.hkr.se

**Abstract**: Based on cyclic simplex codes, a new construction of a family of 2-generator quasi-cyclic two-weight codes is given. New optimal binary quasi-cyclic [195, 8, 96], [210, 8, 104] and [240, 8, 120] codes, good QC ternary [195, 6, 126], [208, 6, 135], [221, 6, 144] codes are thus obtained. Furthermre, binary quasi-cyclic self-complementary codes are also constructed.

## I. INTRODUCTION

A code is said to be quasi-cyclic if every cyclic shift of a codeword by *p* positions results in another codeword [1]. Therefore quasi-cyclic (QC) codes are a generalization of cyclic codes with *p* = 1.

A linear code is called projective if any two of its coordinates are linearly independent, or in other words, if the minimum distance of its dual code is at least three. A code is said to be two-weight if any non-zero codeword has a weight of $w_1$ or $w_2$. Two-weight codes are closely related to strongly regular graphs.

In this paper, a new construction of 2-generator quasi-cyclic (QC) two-weight codes is presented. Some new good QC codes are obtained, and binary self-complementary codes are constructed based on the 2-generator QC codes.

## II. CYCLIC CODES AND QC CODES

A. Cyclic Hamming Codes and Simplex Codes

A q-ary linear [n, k, d] code [2] is a k-dimensional subspace of an n-dimensional vector space over GF(q), with minimum distance d between any two codewords. A code is said to be cyclic if every cyclic shift of a codeword is also a codeword. A cyclic code is described by the polynomial algebra. A cyclic [n, k, d] code has a unique generator polynomial g(x). It is a polynomial with degree of n – k. All codewords of a cyclic code are multiples of g(x) modulo $x^n - 1$.

It is well known that for any integer k, there is a simplex [n, k, d] code with distance $d = q^{k-1}$, where $n = (q^k - 1)/(q - 1)$. It should be noted that simplex codes are equidistance codes where $q^k - 1$ non-zero codewords have weights of $q^{k-1}$.

B. Quasi-Cyclic Codes

A code is said to be quasi-cyclic (QC) if a cyclic shift of any codeword by p positions is still a codeword. Thus a cyclic code is a QC code with *p* = 1. The block length n of a QC code is a multiple of *p*, or $n = m \times p$.

Circulants, or cyclic matrices, are basic components in the generator matrix for a QC code. An $m \times m$ cyclic or circulant matrix is defined as

$$C = \begin{bmatrix} c_0 & c_1 & \cdots & c_{m-1} \\ c_{m-1} & c_0 & \cdots & c_{m-2} \\ c_{m-2} & c_{m-1} & \cdots & c_{m-3} \\ \vdots & \vdots & \cdots & \vdots \\ c_1 & c_2 & \cdots & c_0 \end{bmatrix} \quad (1)$$

and it is uniquely specified by a polynomial formed by the elements of its first row, $c(x) = c_0 + c_1 x + c_2 x^2 + \ldots + c_{m-1} x^{m-1}$, with the least significant coefficient on the left.

A 1-generator QC code has the following form of the generator matrix [3]:

$$G = [\, G_0 \; G_1 \; G_2 \; \ldots \; G_{p-1} \,] \quad (2)$$

where $G_i$, $i = 0, 1, 2, \ldots, p-1$, are circulants of order m. Let $g_0(x), g_1(x), \ldots, g_{p-1}(x)$ are the corresponding defining polynomials.

A 2-generator QC [$m \times p$, $k$] codes has the generator matrix of the following form:

$$G = \begin{bmatrix} G_{00} & G_{01} & \ldots & G_{0,p-1} \\ G_{10} & G_{11} & \ldots & G_{1,p-1} \end{bmatrix} \quad (3)$$

where $G_{ij}$ are circular matrices, for $i = 0$, and 1, $j = 0, 1, \ldots, p-1$.

Similarly, a 3-generator QC [$m \times p$, $k$] codes has the generator matrix of the following form:

$$G = \begin{bmatrix} G_{00} & G_{01} & \ldots & G_{0,p-1} \\ G_{10} & G_{11} & \ldots & G_{1,p-1} \\ G_{20} & G_{21} & \ldots & G_{2,p-1} \end{bmatrix} \quad (4)$$

where $G_{ij}$ are circular matrices, for $i = 0, 1$, and 2, $j = 0, 1, \ldots, p-1$.

## III. CONSTRUCTIONS OF 2-GENERATOR QC TWO-WEIGHT CODES

### A. Two-Weight Codes

A linear code is called projective if any two of its coordinates are linearly independent, or in other words, if the minimum distance of its dual code is at least three. A code is said to be two-weight if any non-zero codeword has a weight of $w_1$ or $w_2$, where $w_1 \neq w_2$. A two weight code is also written as the [n, k; $w_1$, $w_2$] code. Two-weight codes are closely related to strongly regular graphs.

In the survey paper [4], Calderbank and Kantor presented many known families of two-weight codes. Among those families, there is a family of two-weight [n, k; $w_1$, $w_2$] codes over GF(q) noted by SU2, that has the following parameters:

Block length $n = i(q^t - 1)/(q - 1)$

Dimension $k = 2t$

Weights $w_1 = (i - 1) q^{t-1}$, $w_2 = i q^{t-1}$

where $2 \leq i \leq q^t$.

In this section, 2-generator QC two-weight codes with the same parameters as SU2 are constructed from cyclic simplex codes.

### B. Binary 2-Generator QC 2-Weight Codes

Given any positive integer k. If there exist a binary cyclic Hamming [$2^k - 1$, $2^k - k - 1$, 3] codes, then there exist a cyclic simplex [$2^k - 1$, k, $2^{k-1}$] code. Let $g_1(x)$ be the generator polynomial of the simplex code, $C_1$. A binary 2-generator QC two-weight [($2^k - 1$)p, 2k] code can be constructed with the following generator matrix:

$$G = \begin{bmatrix} g_1(x) & g_1(x) & g_1(x) & \ldots & g_1(x) \\ 0 & g_1(x) & x g_1(x) & \ldots & x^{i-2} g_1(x) \end{bmatrix} \quad (5)$$

where $2 \leq i \leq 2^k$, is an integer.

Based on the generator matrix structure, and property of the simplex code, it is obvious that any non-zero codeword has a weight $w_1 = (i-1) 2^{k-1}$, or $w_2 = i 2^{k-1}$. So the 2-generator QC codes defined by (5) are two-weight codes in the family SU2.

Example 1. $n = 7$, $k = 3$. $x^7 - 1 = (x + 1)(x^3 + x + 1)(x^3 + x^2 + 1)$. So a cyclic simplex

[7, 3, 4] code is defined by $g_1(x) = x^4 + x^2 + x + 1$. With the construction, 2-generator QC two-weight [14, 6; 4, 8], [21, 6; 8, 12], [28, 6; 12, 16], [35, 6; 16, 20], [42, 6; 20, 24], [49, 6; 24, 28] and [56, 6; 28, 32] codes are obtained.

Among the QC two-weight codes obtained, some codes are optimal codes, in the sense that they meet the bound [5] on the minimum distance. Table I lists these optimal binary 2-generator QC codes constructed.

Table I OPTIMAL BINARY 2-GENERATOR QC [pm, 2k] CODES

| p | m | k | d | $w_1, w_2$ |
|---|---|---|---|---|
| 3 | 7 | 3 | 8 | 8, 12 |
| 4 | 7 | 3 | 12 | 12, 16 |
| 5 | 7 | 3 | 16 | 16, 20 |
| 6 | 7 | 3 | 20 | 20, 24 |
| 7 | 7 | 3 | 24 | 24, 28 |
| 8 | 7 | 3 | 28 | 28, 32 |
| 10 | 15 | 4 | 72 | 72, 80 |
| 11 | 15 | 4 | 80 | 80, 88 |
| 12 | 15 | 4 | 88 | 88, 96 |
| 13 | 15 | 4 | 96 | 96, 104 |
| 14 | 15 | 4 | 104 | 104, 112 |
| 15 | 15 | 4 | 112 | 112, 120 |
| 16 | 15 | 4 | 120 | 120, 128 |

Among those codes, QC [195, 8, 96], [210, 8, 104] and [240, 8, 120] codes are previously unknown[6].

C. q-ary 2-Generator QC 2-Weight Codes

For any prime power $q$, there exist a *q-ary* cyclic simplex $[(q^k-1)/(q-1), k, q^{k-1}]$ code, if $q-1$ and $k$ are relatively prime. Let $g_1(x)$ be the generator polynomials. Let $m = (q^k-1)/(q-1)$. In the same way as the binary 2-generator QC code construction, we can construct a *q-ary* 2-generator QC two-weight $[m \times p, 2k]$ code with the following generator matrix:

$$G = \begin{bmatrix} g_1(x) & g_1(x) \\ 0 & a_j x^i g_1(x) \end{bmatrix} \quad (6)$$

where $0 \leq i < m$, is an integer, and $a_j$ is any non-zero element in GF(q).

Example 2. n = 13, k = 3. $g_1(x) = x^{10} - x^9 + x^8 - x^6 - x^5 + x^4 + x^3 + x^2 + 1$ defines a cyclic simplex [13, 3, 9] code over GF(3). So 2-generator QC two-weight [26, 6; 9, 18] and [39, 6; 18, 27] codes can be obtained by following generator matrices:

$$G = \begin{bmatrix} g_1(x) & g_1(x) \\ 0 & g_1(x) \end{bmatrix},$$

$$G = \begin{bmatrix} g_1(x) & g_1(x) & g_1(x) \\ 0 & g_1(x) & -g_1(x) \end{bmatrix}$$

Also 2-generator QC two-weight [195, 6, 126], [208, 6, 135], [221, 6, 144] codes over GF(3) are obtained, that reach the lower bound on the minimum distance [5].

IV. CONSTRUCTIONS OF BINARY SELF-COMPLEMENTARY CODES

A binary [n, k, d] code is said to be self-complementary if it has the property that the complementary codeword $(x_1+1, x_2+1, \ldots, x_n+1)$ is also a codeword, for any codeword $(x_1, x_2, \ldots, x_n)$. For a self-complementary [n, k, d] code C, Grey-Rankin bound holds [7]:

$$|C| \leq \frac{8d(n-d)}{n-(n-2d)^2} \quad (7)$$

McGuire [9] has shown that the parameters f a binary linear self-complementary codes meeting the Grey-Rankin bound are

$$[2^{2k-1} - 2^{k-1}, \ 2k+1, \ 2^{2k-2} - 2^{k-1}] \quad (8)$$
$$[2^{2k-1} + 2^{k-1}, \ 2k+1, \ 2^{2k-2}] \quad (9)$$

These self-complementary codes are closely related to quasi-symmetric designs[8, 9]. In [7], Gulliver and Harada investigated 1-generator QC self-complementary [120, 9, 56], [135, 9, 64], [496, 11, 240] and [528, 11, 256] codes. In this section, 3-generator QC self-complementary codes of the parameters as given in (8) and (9) are constructed.

A. $[2^{2k-1} - 2^{k-1}, \ 2k+1, \ 2^{2k-2} - 2^{k-1}]$ Codes

Given a cyclic simplex $[2^k - 1, k, 2^{k-1}]$ code, that is defined by the generator polynomial $g_1(x)$. Choose $i = 2^{k-1}$. Then a 2-generator QC two-weight $[2^{2k-1} - 2^{k-1}, 2k; 2^{2k-2} - 2^{k-1}, 2^{2k-2}]$ code can be constructed by ( ). So, the sum of two non-zero weights is $(2^{2k-2} - 2^{k-1}) + 2^{2k-2} = 2^{2k-1} - 2^{k-1}$, the block length of the code. By extending one more information digit, a 3-generator QC self-complementary $[2^{2k-1} - 2^{k-1}, 2k + 1, 2^{2k-2} - 2^{k-1}]$ Code is obtained by the following generator matrix:

$$G = \begin{bmatrix} g_1(x) & g_1(x) & g_1(x) & g_1(x) \\ 0 & g_1(x) & xg_1(x)... & x^{i-2}g_1(x) \\ 1(x) & 1(x) & 1(x) & 1(x) \end{bmatrix} \quad (10)$$

where $1(x)$ is a vector of all 1's of length $2^k - 1$.

B. $[2^{2k-1} + 2^{k-1}, 2k + 1, 2^{2k-2}]$ Codes

Given a cyclic simplex $[2^k - 1, k, 2^{k-1}]$ code, that is defined by the generator polynomial $g_1(x)$. Choose $i = 2^{k-1} + 1$. Then a 2-generator QC two-weight $[2^{2k-1} + 2^{k-1} - 1, 2k; 2^{2k-2}, 2^{2k-2} + 2^{k-1}]$ code can be constructed by ( ). So, the sum of two non-zero weights is $2^{2k-2} + (2^{2k-2} + 2^{k-1}) = 2^{2k-1} + 2^{k-1}$. By extending one more information digit, and one parity check digit, a 3-generator QC self-complementary $[2^{2k-1} + 2^{k-1}, 2k + 1, 2^{2k-2}]$ Codes is obtained by the following generator matrix:

$$G = \begin{bmatrix} g_1(x) & g_1(x) & g_1(x) & g_1(x) & 0 \\ 0 & g_1(x) & xg_1(x)... & x^{i-2}g_1(x) & 0 \\ 1(x) & 1(x) & 1(x) & 1(x) & 1 \end{bmatrix} \quad (11)$$

where $1(x)$ is a vector of all 1's of length $2^k - 1$.

## V. CONCLUSION

In this paper, a new construction method for a family of two-weight codes is presented. With this construction, some new optimal and good QC codes are obtained, and binary self-complementary codes are constructed by extending the 2-generator QC two-weight codes.